\documentclass[a4paper,12pt]{article}
\usepackage[top=2.5cm, bottom=2.5cm, left=2.5cm, right=2.5cm]{geometry}
\usepackage{authblk, graphicx, amsmath, amssymb, booktabs, multicol, subcaption}

\title{Measurements of sub-nT dynamic magnetic field shielding with soft iron and mu-metal for use in linear colliders}
\author[1, 2]{C. Gohil}
\author[1]{P.\,N. Burrows}
\author[2]{N. Blaskovic Kraljevic\footnote{Present address: European Spallation Source, Lund, Sweden.}}
\author[2]{D. Schulte}
\author[3]{B. Heilig}
\affil[1]{\small John Adams Institute, University of Oxford, Oxford, United Kingdom}
\affil[2]{\small European Organization for Nuclear Research, Geneva, Switzerland}
\affil[3]{\small Mining and Geological Survey of Hungary, Tihany, Hungary}
\date{\small (\today)}

\begin{document}
\maketitle

\begin{abstract}
There is an increasing need to shield beams and accelerator elements from stray magnetic fields. The application of magnetic shielding in linear colliders is discussed. The shielding performance of soft iron and mu-metal is measured for magnetic fields of varying amplitude and frequency. Special attention is given to characterise the shielding performance for very small-amplitude magnetic fields.
\end{abstract}

\section{Introduction}
Magnetic fields can influence the operation of an accelerator. This could be a direct impact on the beam or an influence on accelerator elements. Linear colliders (described below) have an unprecedented sensitivity to external dynamic (stray) magnetic fields.

\subsection{Linear Colliders and Sensitivity to Sub-nT Stray Magnetic Fields} \label{s:linear-colliders}
The luminosity of a linear $e^+ e^-$ collider is \cite{beam-beam-effects}
\begin{equation}
\mathcal{L} = \frac{N^2 f_\text{rep} n_b}{4\pi \sigma^*_x \sigma^*_y} H_D,
\end{equation}
where $N$ is the bunch population, $f_\text{rep}$ is the repetition frequency, $n_b$ is the number of bunches, $\sigma^*_x$ ($\sigma^*_y$) is the horizontal (vertical) beam size at the interaction point and $H_D$ is a luminosity enhancement factor due to the electromagnetic interaction between the colliding bunches. To achieve a large luminosity extremely small vertical beam sizes are targeted. These beam sizes can only be realised with ultra-low emittance beams. Stray magnetic fields lead to emittance growth and a beam-beam offset at collision, which reduces luminosity. A more comprehensive description of the impact of stray magnetic fields can be found in \cite{thesis}.

Simulation studies of the Compact Linear Collider (CLIC) \cite{clic-cdr, clic-pip} show stray magnetic field amplitudes on the order of 0.1\,nT for the 380 GeV stage~\cite{thesis, sf-tolerances} and amplitudes of 1\,nT for the 3\,TeV stage \cite{snuverink} can lead to a 2\% luminosity loss. This leads to a worst-case tolerance of 0.1\,nT for stray magnetic fields in CLIC. For the International Linear Collider (ILC), a sensitivity to stray magnetic field amplitudes of 1\,nT has been reported~\cite{ilc-tdr}. Measurements of the ambient magnetic field in accelerator environments have shown RMS amplitudes of up to 100\,nT over the frequency range 0.1-3\,kHz~\cite{thesis, sf-measurements}. Therefore to avoid performance loss, a mitigation system will be essential.

There are two options for such a system: an active compensation device or a passive shielding system. An active compensation device would measure the magnetic field and use a set of coils to compensated it. Such a device was demonstrated at an accelerator facility in~\cite{active-compensation}. This system stabilised a magnetic field to fluctuations of less than 10\,nT. However, an active compensation device relies on accurately measuring the magnetic field. Measuring magnetic field fluctuations of 0.1\,nT is challenging with current commercially available magnetometers~\cite{ripka}. Therefore, a passive shielding system is preferred. A shielding factor of approximately $10^3$ is required to reduce a 100\,nT stray magnetic field to the level of 0.1\,nT.

\subsection{Magnetic Shielding in Linear Colliders}
The most common use for magnetic shields in linear colliders is for superconducting radio-frequency (SRF) cavities, which are used in the ILC~\cite{ilc-tdr}. SRF cavities must be cooled down to superconducting temperatures to operate, usually 2\,K. If magnetic flux is trapped inside the cavity walls during the cool down the quality factor of the cavity is reduced~\cite{flux-trapping}. A magnetic shield is used to prevent magnetic flux trapping. Studies of potential magnetic shields for ILC SRF cavities are presented in~\cite{kek1, kek2, kek3}.

In the above application, the magnetic shield is used to shield static magnetic fields. In this work, we look at the use of magnetic shields to shield the beam from dynamic magnetic fields. In particular, low-frequency small-amplitude magnetic fields.

\subsection{Shielding Mechanisms and Magnetic Permeability}
\begin{figure}[!htb]
\centering
\includegraphics[width=0.25\linewidth]{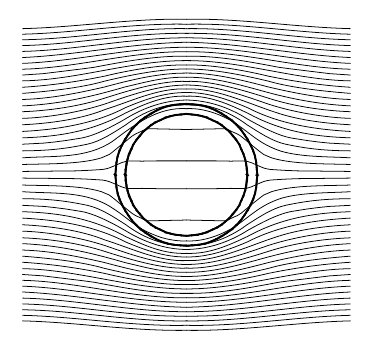}
\hfil
\includegraphics[width=0.25\linewidth]{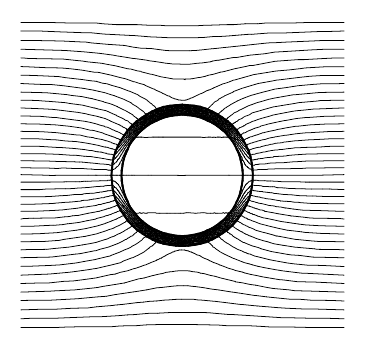}
\caption{\small Cylindrical shields subject to a uniform magnetic field~\cite{celozzi}. Left: flux shunting and right: eddy-current cancellation.}
\label{f:shielding-mechanisms}
\end{figure}

An overview of magnetic shielding is given in~\cite{sumner1987convectional}. There are two magnetic shielding mechanisms, which are shown in Figure \ref{f:shielding-mechanisms}. On the left is the flux-shunting mechanism, which is effective for shielding static and low-frequency magnetic fields, and on the right is eddy-current cancellation, which is only effective for high-frequency magnetic fields. In this paper, we study the shielding of low-frequency magnetic fields, therefore the flux-shunting mechanism is of interest.

The flux-shunting mechanism relies on the material possessing a large permeability to draw the magnetic field away from the shielded region. Ferromagnetic materials~\cite{jiles} are commonly used for this purpose. The permeability of a ferromagnetic material that is exposed to a dynamic magnetic field is given by
\begin{equation}
\mu(H) = \frac{B(H)}{H},
\label{e:incremental-permeability}
\end{equation}
where $H$ is the amplitude of the magnetic field variations and $B(H)$ is the amplitude of the magnetic induction. The permeability is independent of a static offset provided the material is not close to saturation.

\subsection{Shielding Small-Amplitude Magnetic Fields}
The response of a material to a magnetic field with an amplitude much smaller than the coercive field of the material is governed by Rayleigh's law, which states the amplitude of the magnetic induction is given by~\cite{jiles}
\begin{equation}
B(H) = \mu_i H + \nu H^2,
\end{equation}
where $\mu_i$ is the initial permeability and $\nu$ is Rayleigh's constant. The permeability in the Rayleigh region is given by
\begin{equation}
\mu(H) = \mu_i + \nu H.
\label{e:rayleigh-permeability}
\end{equation}
In order to effectively shield small-amplitude magnetic fields, the material must possess a sufficiently high initial permeability.

As discussed in Sec.\,\ref{s:linear-colliders}, linear colliders have stray magnetic field tolerances as small as 0.1\,nT. A magnetic shield is an attractive system to use if it is effective at these field levels. It is important to measure the permeability as a function of external magnetic field amplitude, i.e. the parameters $\mu_i$ and $\nu$ in Eq.\,\eqref{e:rayleigh-permeability}, in order to validate whether a shield would be effective at a particular field amplitude.

In this paper, we will characterise the permeability and shielding performance of two ferromagnetic materials (discussed in the next section) at the field levels of interest to linear colliders to access whether these materials will be effective for magnetic shielding in linear colliders. To the writer's knowledge these measurements cannot be found in existing literature in the accelerator physics community. Shielding low fields is also of interest in the design of SQUID systems. Work that looks at magnetic shielding for SQUIDs can be found in~\cite{squid1, squid2, squid3, squid4}.

\section{Measurements}
Two ferromagnetic materials were characterised: soft iron and a nickel-iron alloy known as mu-metal. The magnetic shielding performance is measured with a transfer function, which is described below.

\subsection{Transfer Functions}
Considering a magnetic shield exposed to the time-varying magnetic field $H_e e^{j2\pi f t}$, where $f$ is the frequency, $t$ is the time, $H_e$ is the external magnetic field amplitude and $j=\sqrt{-1}$, the magnetic field in the shielded region is $H_i e^{j (2\pi f t - \phi)}$, where $\phi$ is a phase shift introduced by the shield and $H_i$ is the internal magnetic field amplitude. The transfer function of the magnetic shield is given by
\begin{equation}
T(f) = \frac{H_i e^{-j\phi}}{H_e},
\end{equation}
The absolute value of $T(f)$ is known as the amplitude response and the phase of $T(f)$ is known as the phase response.

For simple geometries, such as an infinitely long cylinder, analytical solutions to Maxwell's equations exist for the propagation of electromagnetic waves through magnetic shields. A method for calculating the shielding factor of cylindrical shields is described in~\cite{hoburg}.

\subsection{Methodology}
A cylinder of inner diameter 5\,cm, thickness 1\,mm and length 50\,cm was formed from soft iron and another cylinder with the same dimensions were formed from mu-metal. Both cylinders were annealed in their final form. The advertised magnetic properties of each material (provided by the supplier) are summarised in Table~\ref{t:material-specifications}.

\begin{table}[!htb]
\centering
\begin{tabular}{l c c}
\toprule
\textbf{Property} & \textbf{Soft Iron} & \textbf{Mu-Metal} \\
\midrule
Initial permeability & 300-500 & 50,000 \\
Maximum relative permeability & 3,500-8,000 & 250,000 \\
Magnetic induction at saturation & 2.15 T & 0.74 T \\
\bottomrule
\end{tabular}
\caption{\small Advertised specifications of each material~\cite{ak-steel, magnetic-shield-corporation}.}
\label{t:material-specifications}
\end{table}

A similar methodology to~\cite{fermilab1} was used into this work to measure the shielding performance of the soft iron and mu-metal cylinder. A three-axis Bartington Mag-13 fluxgate magnetometer \cite{bartington-mag-13} was used in the measurements. Dedicated measurements were performed to characterise the sensor, these are described in~\cite{thesis}. The noise level of this sensor is low enough to measure magnetic field amplitudes of less than 0.1 nT. In this work, a set of Helmholtz coils~\cite{ripka} was used to generate a magnetic field excitation at a precise frequency and amplitude. A Mag-13 sensor was placed at the centre of the Helmholtz coils.

The magnetic field $H(t)$ was measured with and without a shield surrounding the sensor. In both measurements the current in the Helmholtz coils $I(t)$ was simultaneously recorded. A transfer function that relates the current in the Helmholtz coil to the magnetic field measured by the sensor was calculated:
\begin{equation}
T_{IH}(f) = \frac{P_{IH}(f)}{P_{II}(f)},
\end{equation}
where $P_{IH}(f)$ is the cross power spectral density of $I(t)$ and $H(t)$ and $P_{II}(f)$ is the power spectral density of $I$(t). The transfer function for the shield was calculated as
\begin{equation}
T(f) = \frac{T_{IH,\text{sh}}(f)}{T_{IH,\text{no sh}}(f)},
\end{equation}
where $T_{IH,\text{sh}}(f)$ is the transfer function measured with the shield and $T_{IH,\text{no sh}}$ is the transfer function measured without the shield.

\subsection{Soft Iron}
\begin{figure}[!htb]
\centering
\includegraphics[width=0.49\textwidth]{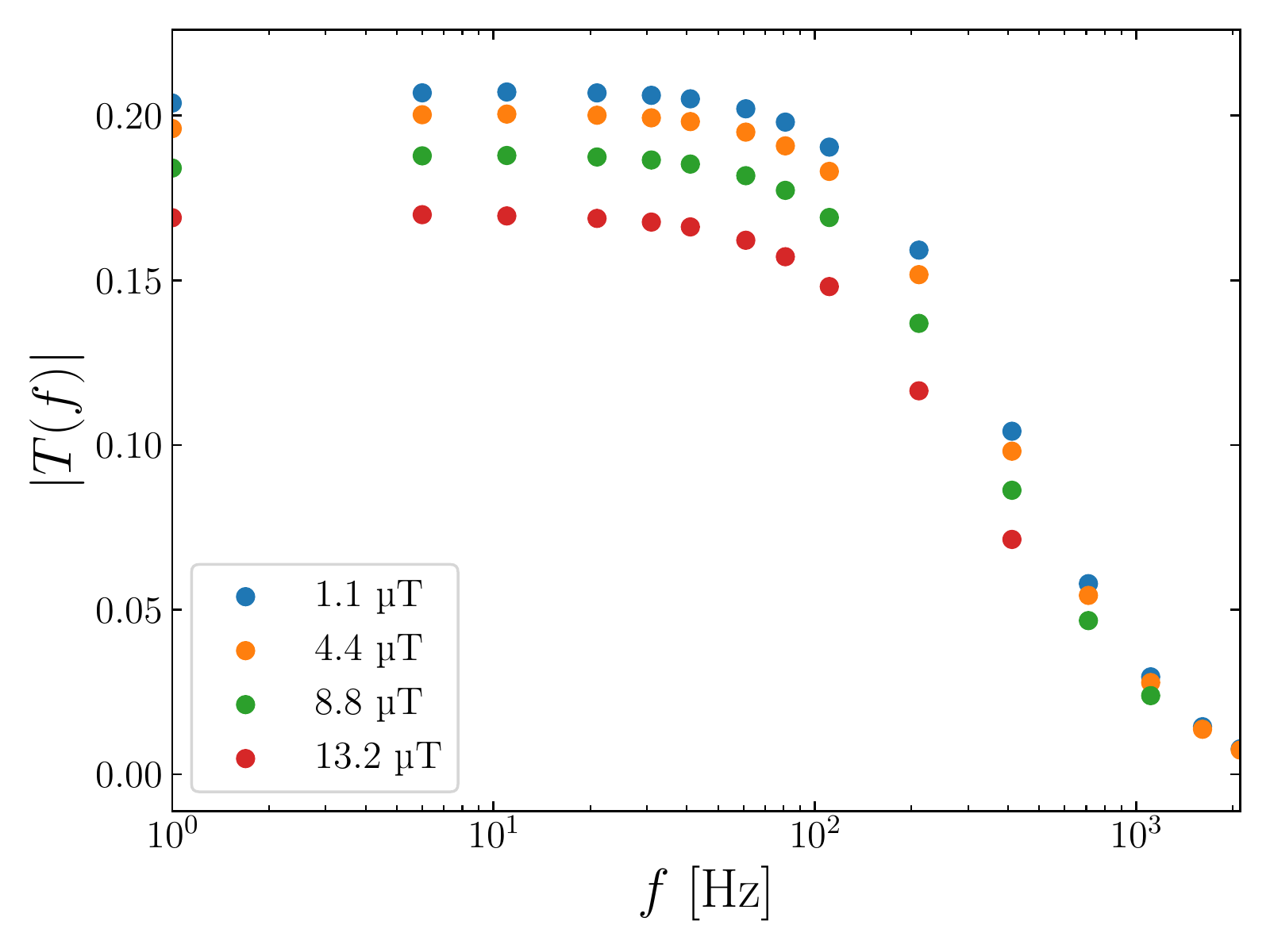}
\includegraphics[width=0.49\textwidth]{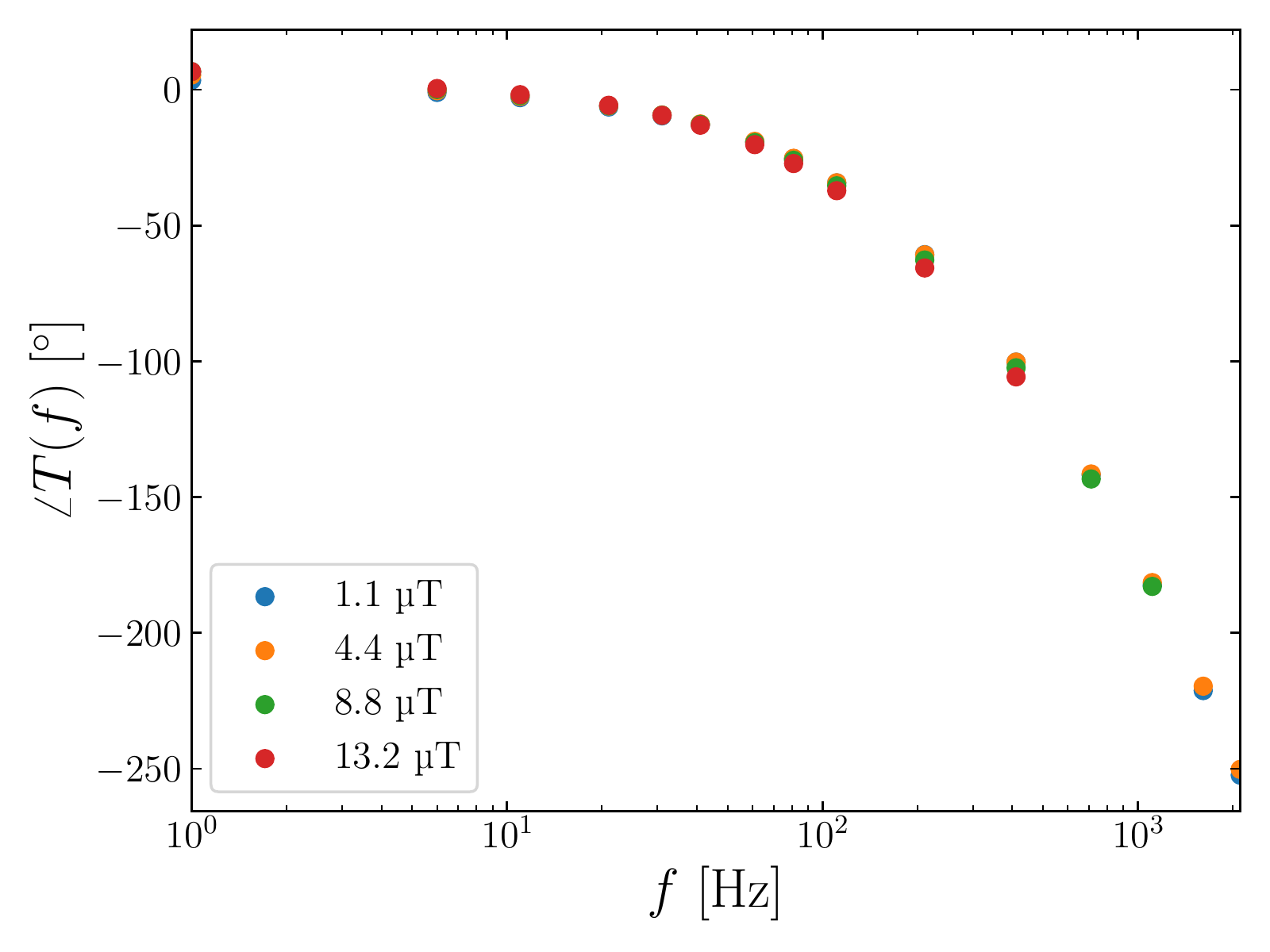}
\caption{\small Transfer function of the iron cylinder for different external magnetic field amplitudes. Left: amplitude response $|T(f)|$ vs frequency $f$. Right: phase response $\angle T(f)$ vs frequency $f$. Error bars are too small to be seen.}
\label{f:iron-tf}
\end{figure}

The transfer function of a high purity (99.9\%) iron cylinder was measured with different external magnetic field amplitudes. The transfer functions are shown in Figure \ref{f:iron-tf}. There is a clear dependence on the external magnetic field amplitude, where the shielding improves with the amplitude. The phase response of the iron cylinder is independent of the external magnetic field amplitude.

\begin{figure}[!htb]
\centering
\includegraphics[width=0.55\linewidth]{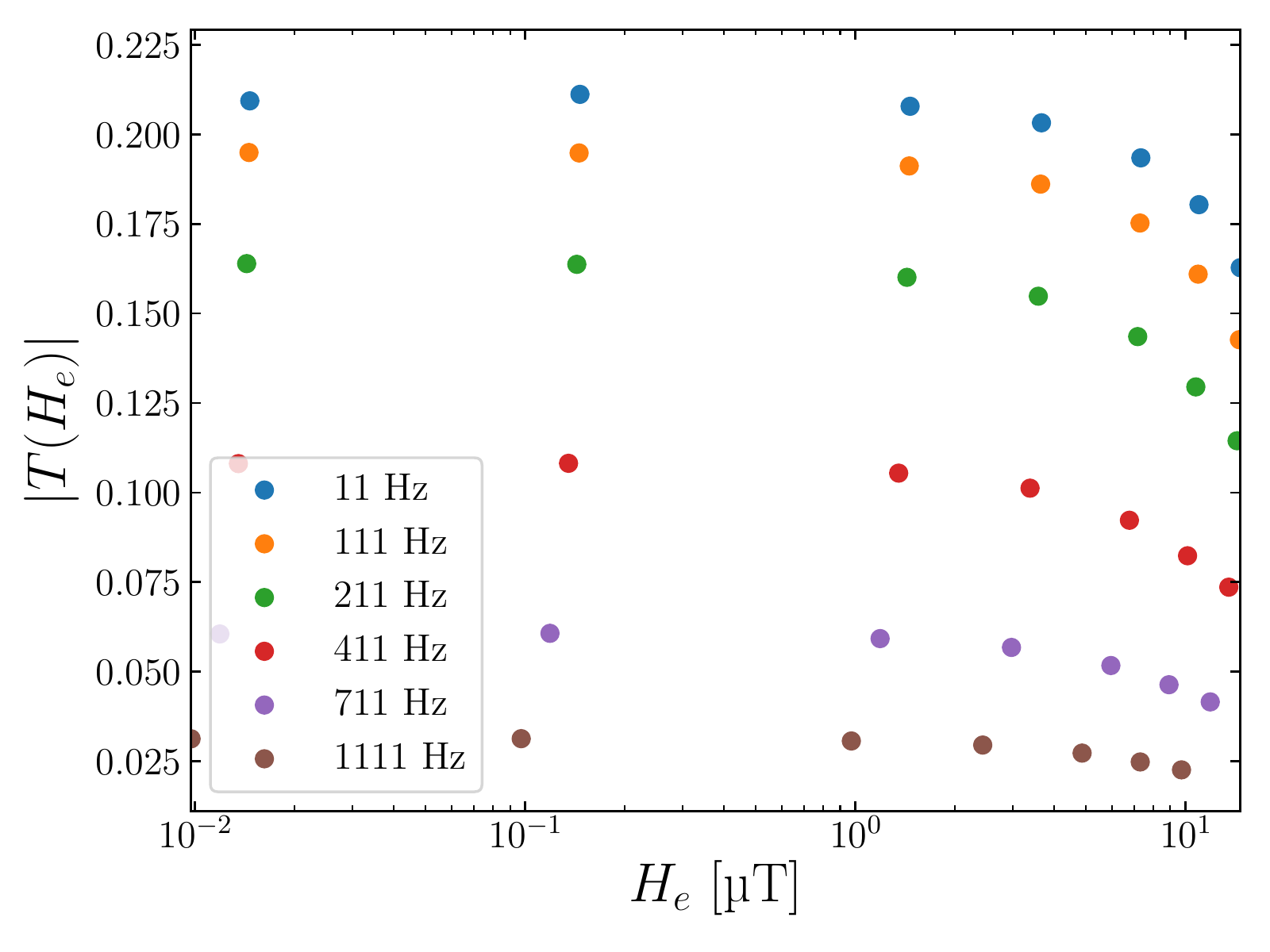}
\caption{\small Amplitude response of the soft iron cylinder $|T(H_e)|$ vs external magnetic field amplitude $H_e$ for different frequencies. Error bars too small to be seen.}
\label{f:iron-linearity}
\end{figure}

\begin{figure}[!htb]
\centering
\includegraphics[width=0.55\linewidth]{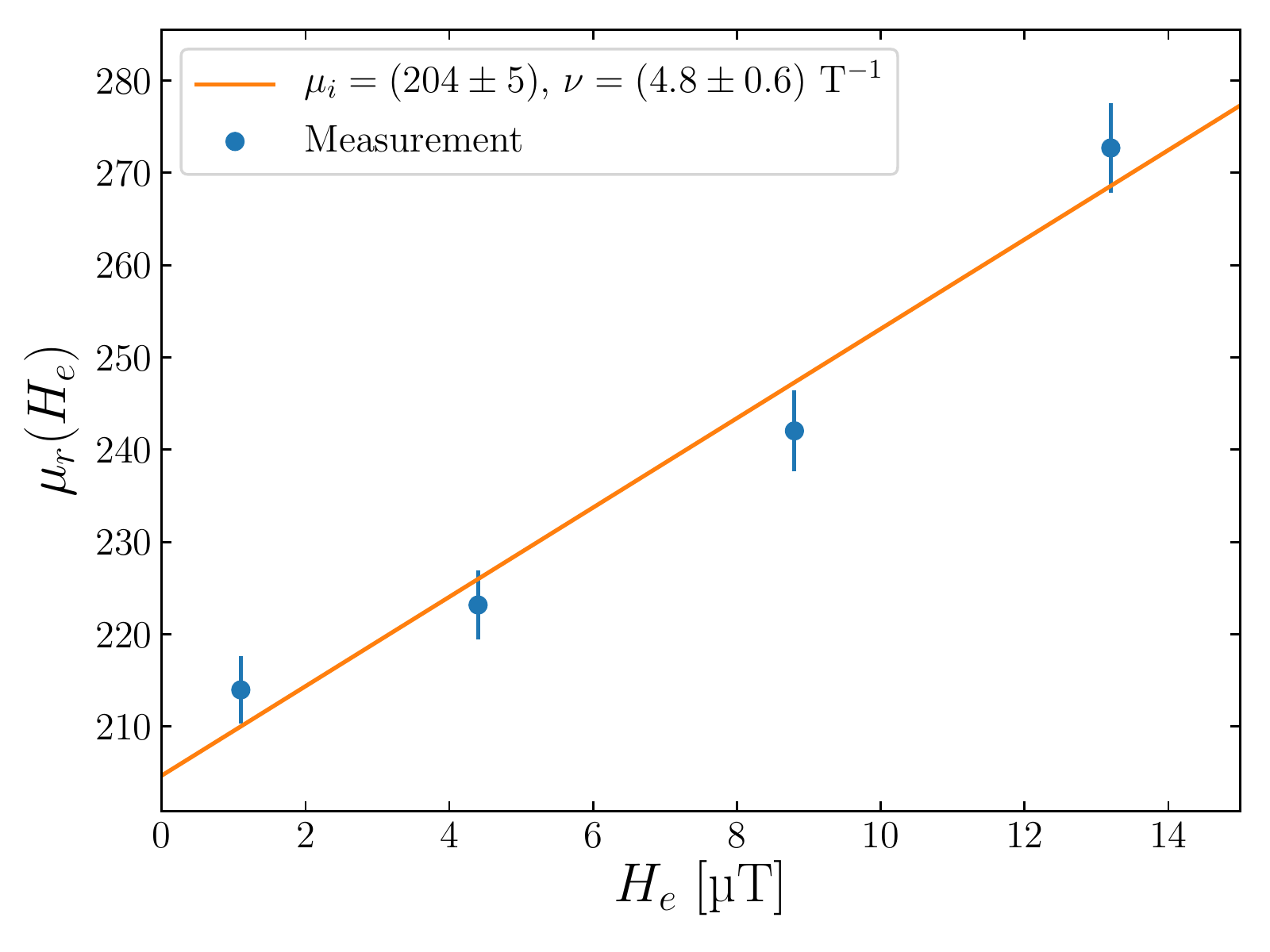}
\caption{\small Relative permeability of the soft iron cylinder $\mu_r(H_e)$ vs external magnetic field amplitude $H_e$: measurement (blue) and a straight line fit (orange). The errors bars were derived from fitting the model described in~\cite{hoburg} to the transfer functions in Fig.\,\ref{f:iron-tf}.}
\label{f:iron-permeability}
\end{figure}

Figure \ref{f:iron-linearity} shows the measured amplitude response as a function of external magnetic field. It is clear the amplitude response tends to a constant as the external magnetic field is decreased. There is a significant reduction in the amplitude response as the external magnetic field is increased, e.g. for the measurement at 11\,Hz, the amplitude response decreases by roughly 25\% over the range of the measurement. This highlights that the impact of the external magnetic field amplitude on the shielding performance can be significant.

The model described in \cite{hoburg} can be used to fit a permeability to the transfer function for each amplitude. Figure \ref{f:iron-permeability} shows the relative permeability as a function of external magnetic field amplitude. The initial permeability is extrapolated by fitting a straight line to the relative permeability. An initial permeability of $\mu_i = (204\pm5)$ was measured for this iron cylinder, which is somewhat below the advertised value of 300-500. There is a significant drop in the permeability of roughly one third over the range of the measurement. This reduction in permeability reflects the increase in the amplitude response in Fig.\,\ref{f:iron-linearity} for lower amplitudes.

\subsection{Mu-Metal}
In the previous section we showed that the permeability of soft iron is significantly altered by the external magnetic field amplitude. In this section we evaluate the sensitivity of the permeability of mu-metal to the external magnetic field amplitude. The chemical composition of the mu-metal used was 80\% Ni, 15\% Fe, 4.5\% Mo, 0.4\% Mn and 0.1\% Si. 

\begin{figure}[!htb]
\centering
\includegraphics[width=0.49\textwidth]{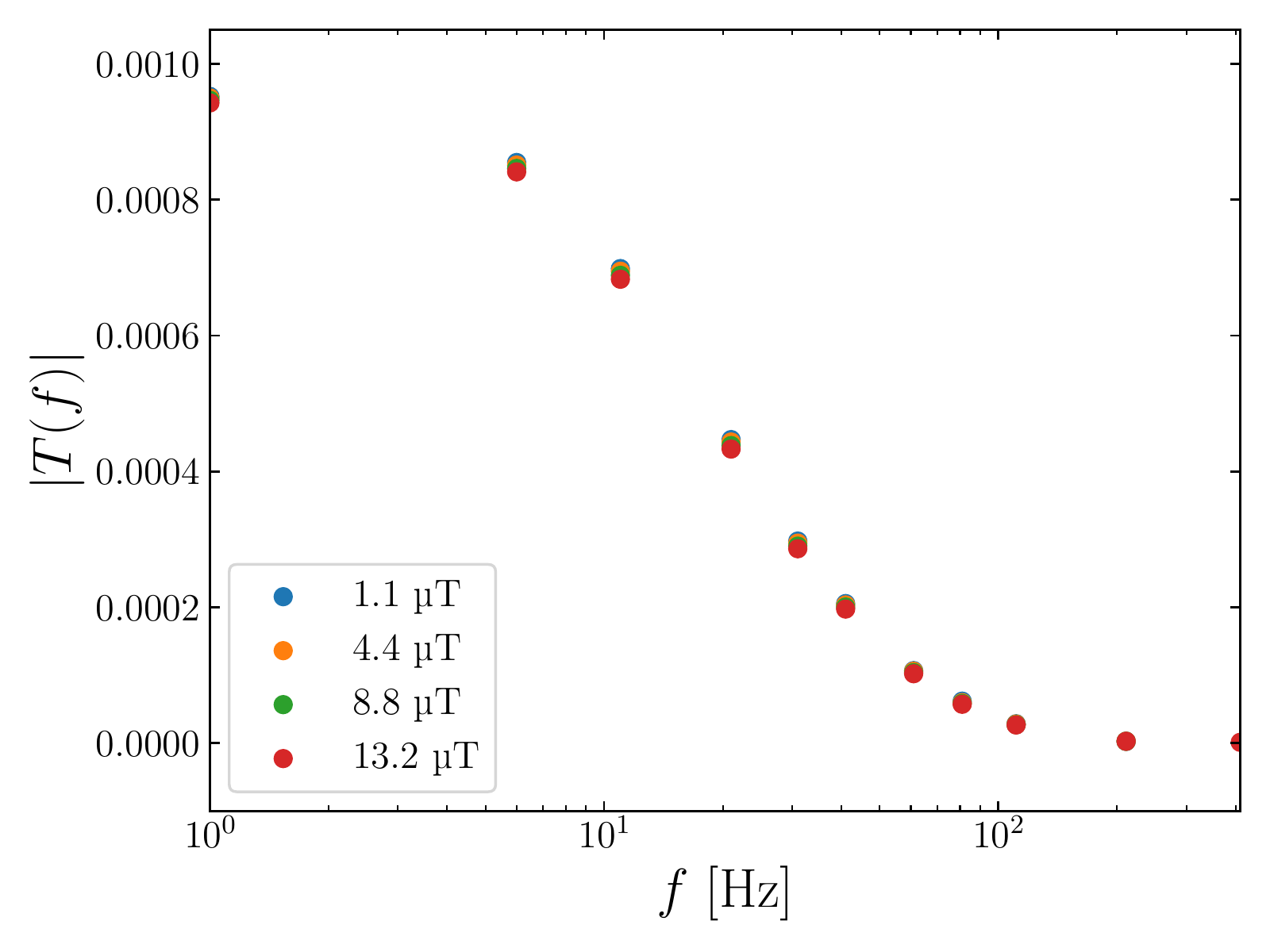}
\includegraphics[width=0.49\textwidth]{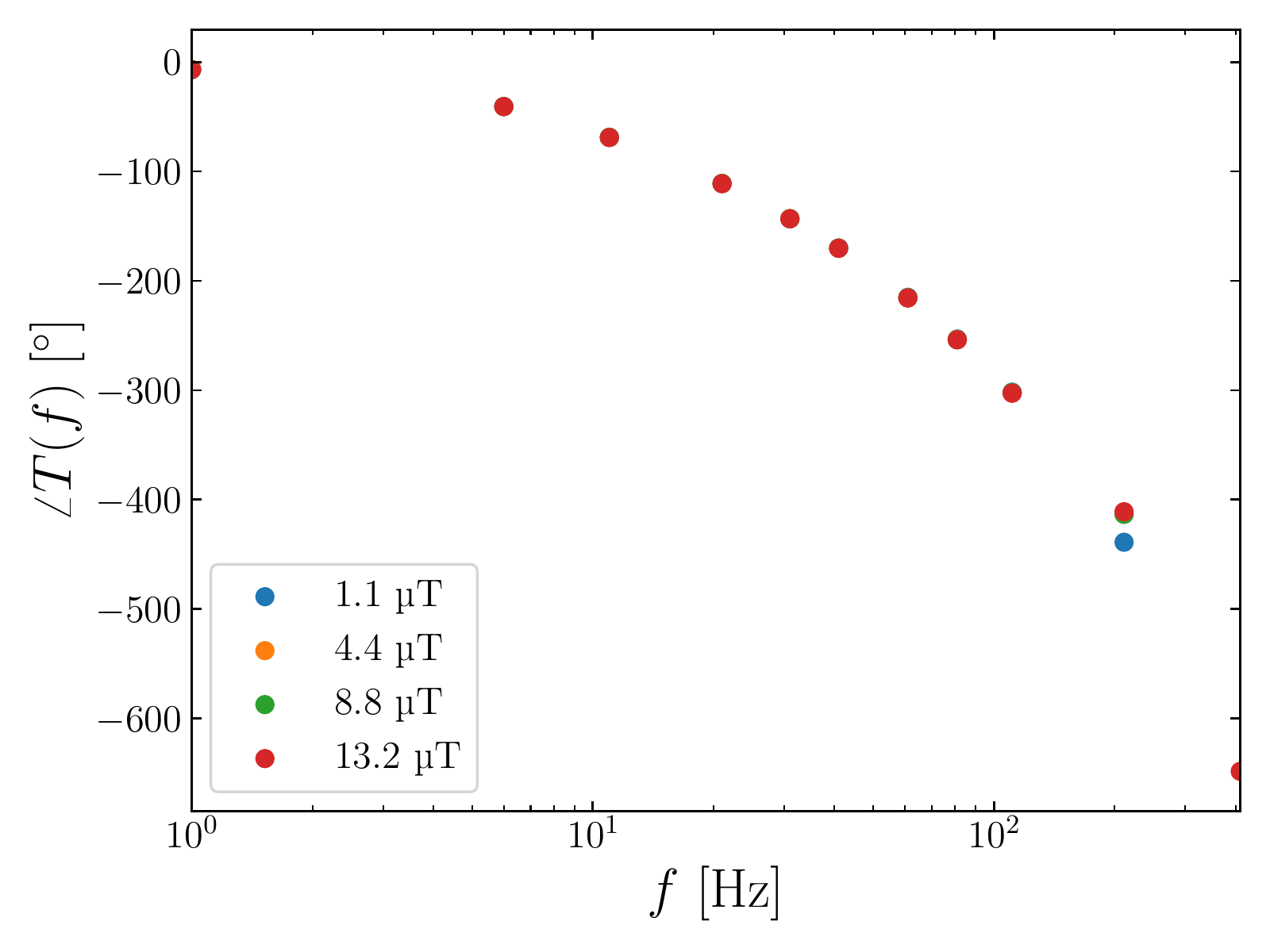}
\caption{\small Transfer function of the mu-metal cylinder for different external magnetic field amplitudes. Left: amplitude response $|T(f)|$ vs frequency $f$. Right: phase response $\angle T(f)$ vs frequency $f$. Error bars are too small to be seen.}
\label{f:mu-metal-tf}
\end{figure}

The transfer function of the mu-metal cylinder measured with different external magnetic field amplitudes is shown in Figure \ref{f:mu-metal-tf}. The transfer function for mu-metal is less sensitive to the external magnetic field amplitude than the soft iron. The Rayleigh constant $\nu$ is an order of magnitude greater for the mu-metal compared to the soft iron.

\begin{figure}[!htb]
\centering
\includegraphics[width=0.55\linewidth]{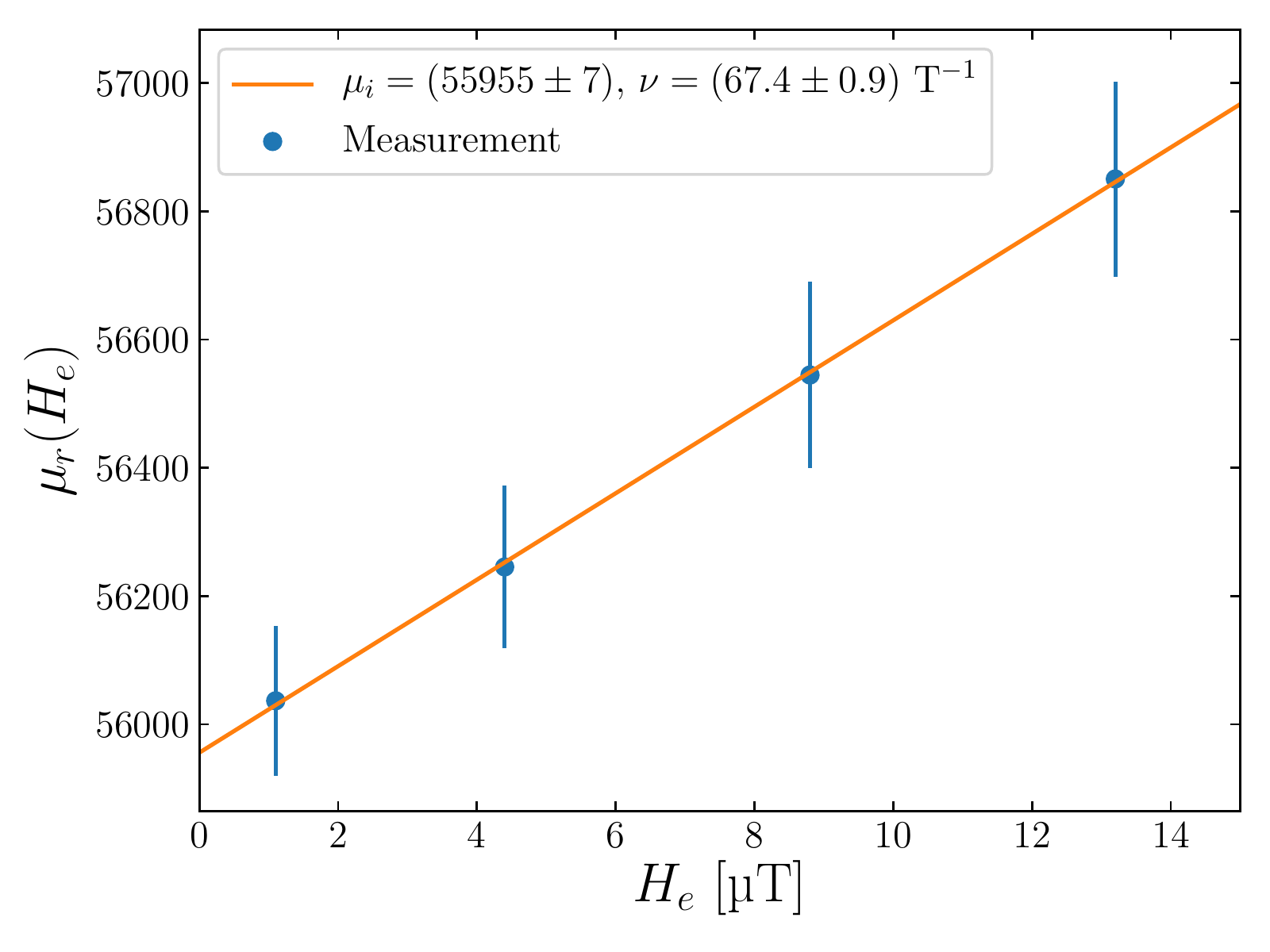}
\caption{\small Relative permeability of the mu-metal cylinder $\mu_r(H_e)$ vs external magnetic field amplitude $H_e$: measurement (blue) and straight line fit (orange). The errors bars were derived from fitting the model described in~\cite{hoburg} to the transfer functions in Fig.\,\ref{f:mu-metal-tf}}
\label{f:mu-metal-permeability}
\end{figure}

Figure \ref{f:mu-metal-permeability} shows the permeability fitted to each transfer function in Figure \ref{f:mu-metal-tf}. There is a clear linear relationship between the permeability and external magnetic field amplitude. It is also clear that the permeability is reduced at for lower external magnetic field amplitudes. This measurement shows that the relative change in permeability over the range measured is much smaller for the mu-metal compared to the soft iron and that the dependence of permeability on external magnetic field amplitude is small. The initial permeability of the mu-metal cylinder is $\mu_i=(55,955\pm7)$, which is above the advertised value of $\mu_i=50,000$. The smaller sensitivity to external magnetic field amplitude for mu-metal makes it an attractive choice to shield low-amplitude fields. The ability of mu-metal to shield to the 0.1\,nT level needed for CLIC is discussed in the next section.

\subsubsection{Shielding to Sub-nT Magnetic Fields}
\begin{figure}[!htb]
\centering
\includegraphics[width=0.55\linewidth]{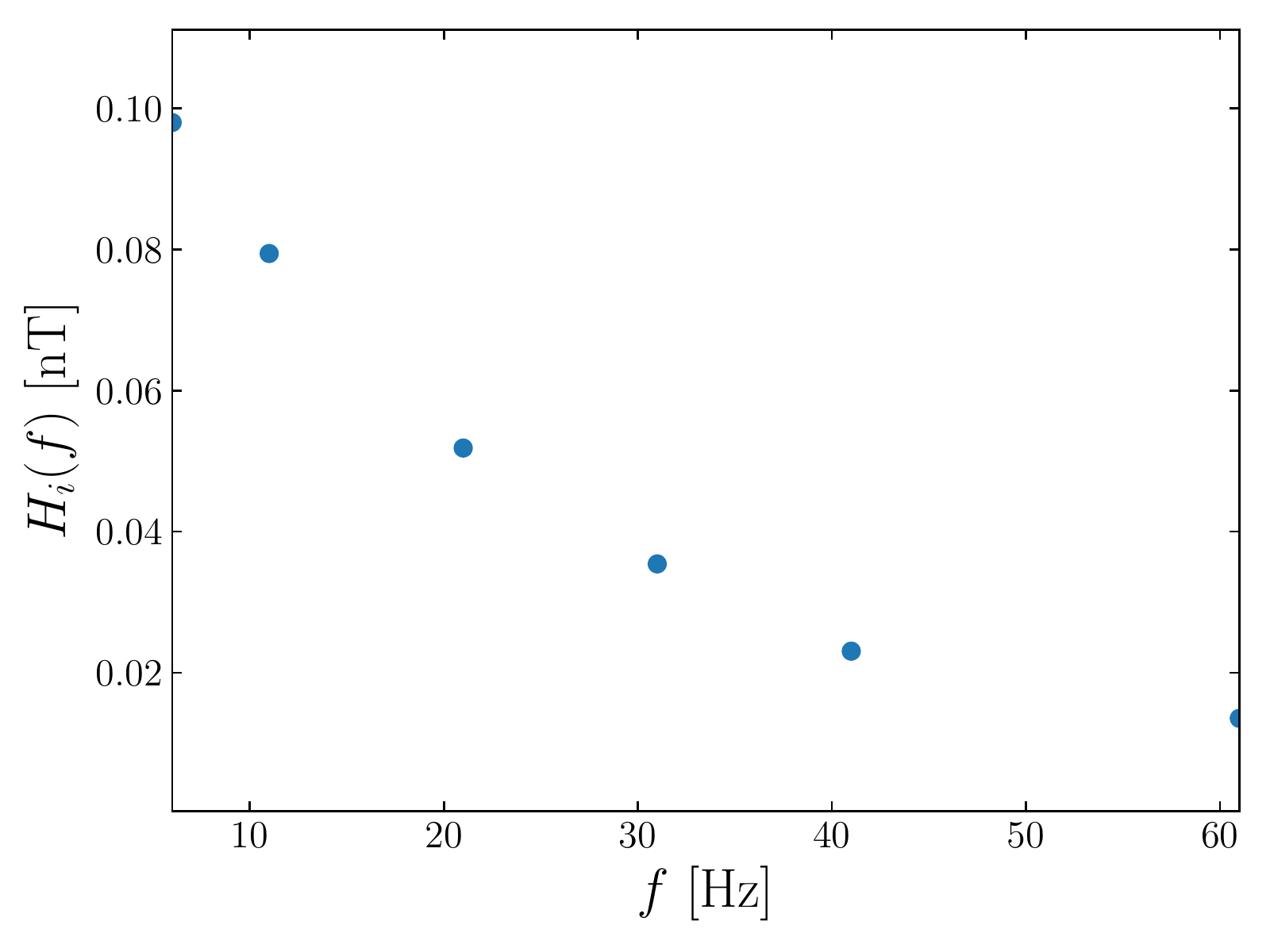}
\caption{\small Internal magnetic field amplitude $H_i(f)$ of the mu-metal cylinder with an external magnetic field amplitude of 1.1 $\mu$T vs frequency $f$.}
\label{f:mu-metal-low-field}
\end{figure}

Realising a sub-nT internal magnetic field requires a very effective magnetic shield with a sufficiently high initial permeability. A 0.1\,nT internal magnetic field amplitude can be demonstrated with the mu-metal shield using an external magnetic field amplitude of 1.1\,$\mu$T. This is shown in Figure \ref{f:mu-metal-low-field}.

The expected amplitude of stray magnetic fields in accelerator environments is up to 100\,nT~\cite{thesis, sf-measurements}, which is an order of magnitude less than the excitation used in the measurement shown in Figure \ref{f:mu-metal-low-field}. At a lower field level, the permeability of the mu-metal will be reduced. However, Figure \ref{f:mu-metal-permeability} shows the effect on the permeability of reducing the external field amplitude from 10\,$\mu$T to 1\,$\mu$T is approximately 1\%. Extrapolating the linear relationship between the field level and permeability, the reduction in permeability from the field level in Figure \ref{f:mu-metal-low-field} to 100\,nT should also be approximately 1\%. Therefore, we can assume the transfer function is roughly the same and we can be confident that the stray field amplitude inside a mu-metal shield will be less than 0.1\,nT for external amplitudes of 100\,nT. With this measurement we can confirm a mu-metal shield can be used to mitigate stray magnetic fields to the level of 0.1\,nT in future linear colliders.

\subsubsection{Mu-Metal Foils} \label{s:mu-metal-foils}
Mu-metal is also available in thin foils, typically of thicknesses 0.1-0.5 mm. These foils are annealed and advertised as retaining their magnetic properties after slight deformation.

\begin{figure}[!htb]
\centering
\includegraphics[width=0.49\textwidth]{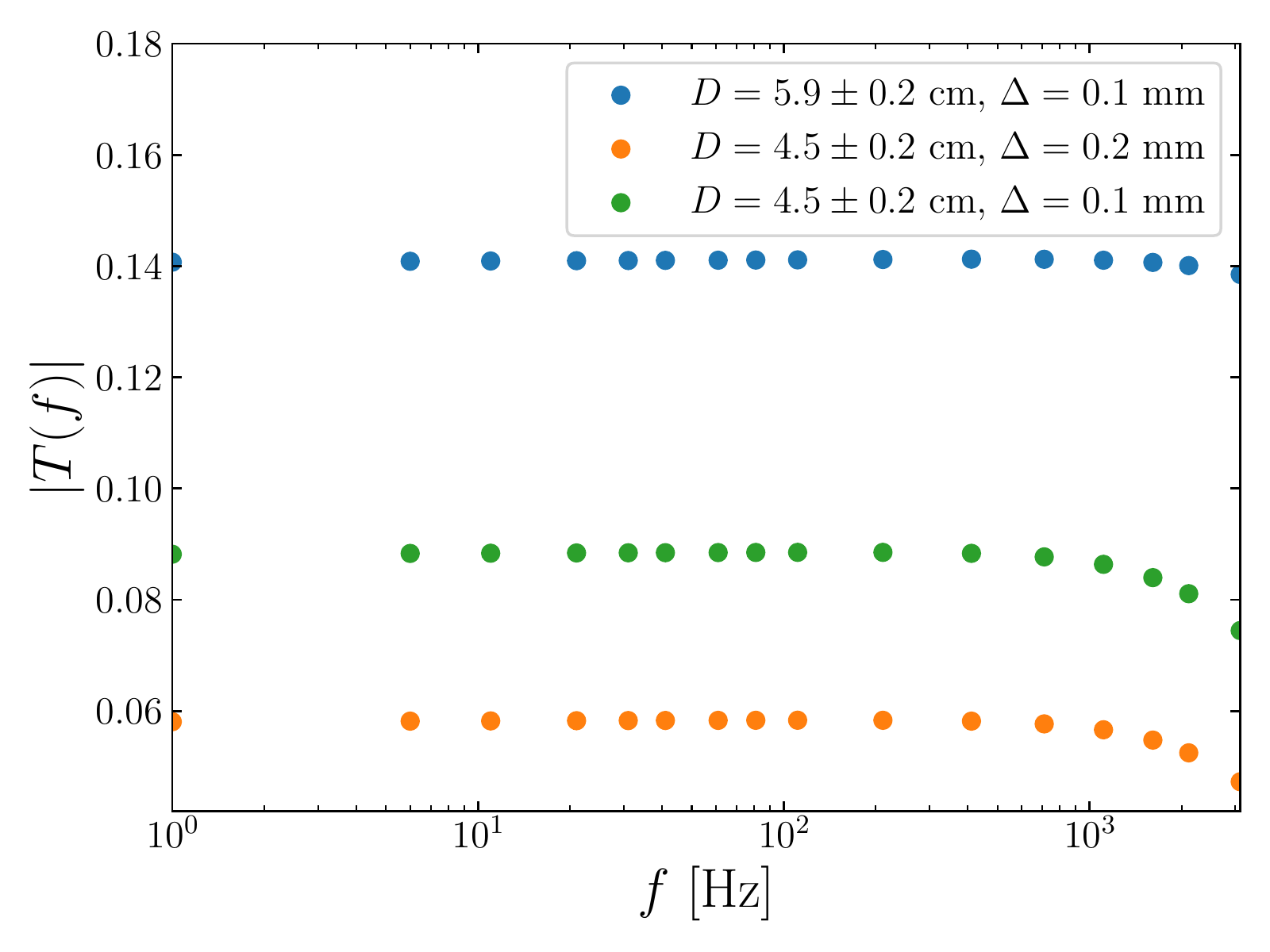}
\includegraphics[width=0.49\textwidth]{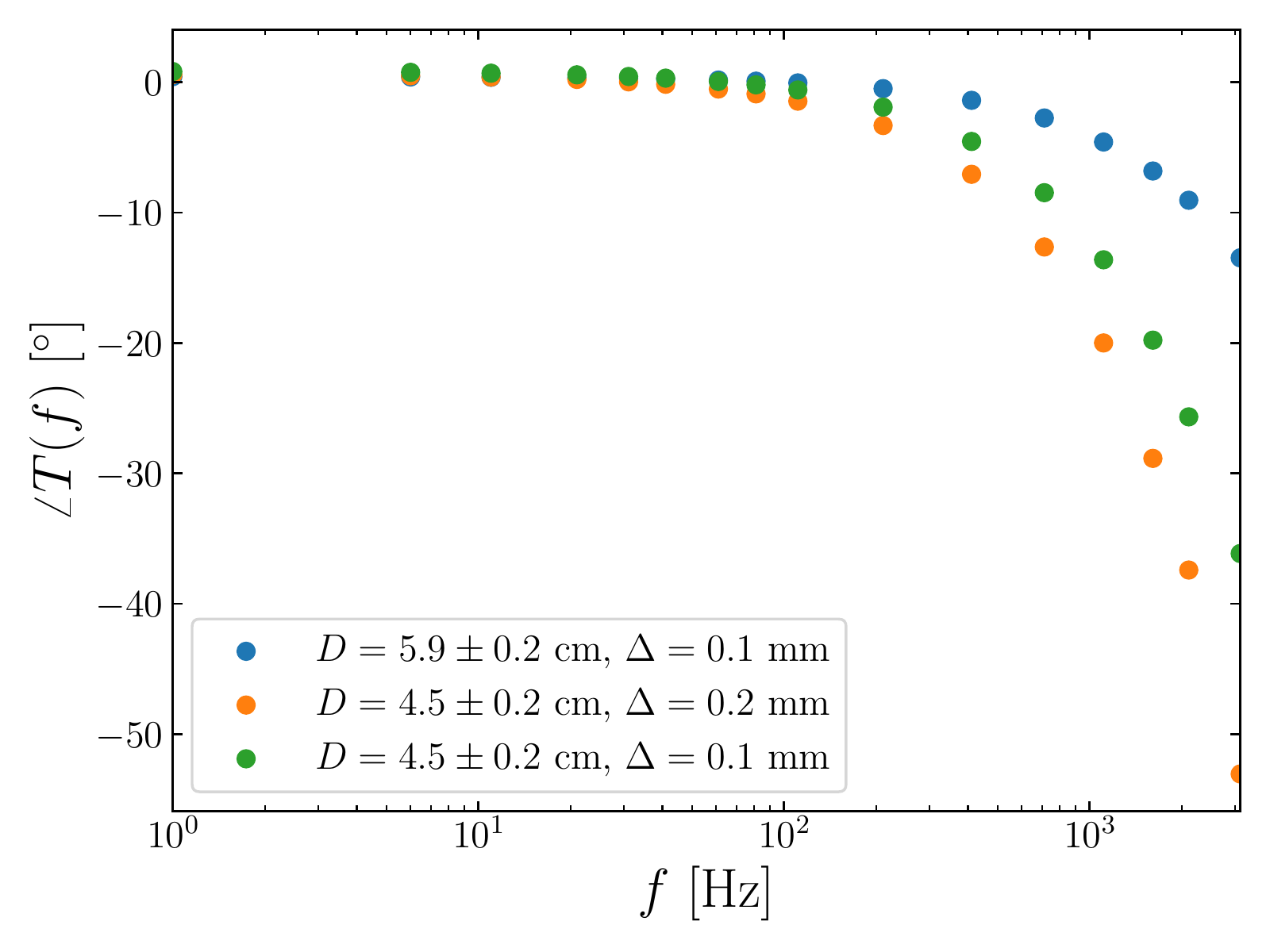}
\caption{\small Transfer function of three shields formed from a mu-metal foil. Left: amplitude response $|T(f)|$ vs frequency $f$. Right: phase response $\angle T(f)$ vs frequency $f$. An external magnetic field amplitude of  1.1 $\mu$T was used. The 0.2 mm thick shield was formed with two layers of foil. Error bars are too small to be seen.}
\label{f:mu-metal-foil-tf}
\end{figure}

A set of three cylindrical shields of varying diameter $D$ and thickness $\Delta$ were formed from a mu-metal foil. These shields were not re-annealed in their final form. The foil had the same chemical composition as the mu-metal cylinder discussed in the previous section. Figure \ref{f:mu-metal-foil-tf} shows the transfer function of each shield formed from the mu-metal foil.

\begin{table}[!htb]
\centering
\begin{tabular}{c c c}
\toprule
\textbf{Diameter,} $\boldsymbol{D}$ \textbf{[cm]} & \textbf{Thickness,} $\boldsymbol{\Delta}$ \textbf{[mm]} & \textbf{Relative Permeability,} $\boldsymbol{\mu_r}$ \ \\
\midrule
$5.9\pm0.2$ & 0.1 & $3,670\pm2$ \\
$4.5\pm0.2$ & 0.2 & $3,602\pm2$ \\
$4.5\pm0.2$ & 0.1 & $4,660\pm50$ \\
\bottomrule
\end{tabular}
\caption{\small Measured relative permeability of three shields formed from a mu-metal foil.}
\label{t:mu-metal-foil-permeability}
\end{table}

Table \ref{t:mu-metal-foil-permeability} shows the permeability fitted to each amplitude response. The foils have a relative permeability of less than 5,000, which is very poor for mu-metal. It is likely that the permeability was damaged by deforming the cylinder when rolling the mu-metal foil to produce the shield, this is discussed in Sec.\,\ref{s:performance-stress}.

A simple model for the shielding factor of a mu-metal shield is presented in~\cite{dubbers}. For a single layer, the amplitude response is given by
\begin{equation}
T = \frac{D}{\mu_r \Delta}.
\label{e:mu-metal-simple-tf}
\end{equation}
This model does not include shielding via the eddy-current cancellation mechanism. For small external magnetic field amplitudes, the relative permeability in Eq.\,\eqref{e:mu-metal-simple-tf} is replaced with the initial permeability. The measured amplitude response for the different mu-metal foils is roughly consistent with Eq.\,\eqref{e:mu-metal-simple-tf}.

\section{Discussion}
Magnetic shields have already been used at different large-scale accelerator facilities, e.g. Brookhaven National Laboratory~\cite{bnl1, bnl2} and Fermilab~\cite{fermilab1, fermilab2}. This section describes various considerations for using the above materials to shield magnetic fields in linear colliders and the factors that affect performance.

\subsection{Beam Pipes}
A beam pipe is used to contain the vacuum in an accelerator. In linear colliders, they typically consist of a few millimetres of steel and a 10-100\,$\mu$m inner copper coating to mitigate long-range wakefields.

The impact of stray magnetic fields can be mitigated by preventing them from reaching the beam. This can be achieved by surrounding the beam with a shield or surrounding the sources with a shield. The beam pipe is usually the closest component to the beam. Therefore, surrounding the beam pipe is the safest option because it prevents stray fields from all external sources reaching the beam. To shield the sources, they must first be identified and the feasibility of surrounding them with a shield must be studied.

Mu-metal is a good candidate material for a magnetic shield. Mu-metal foils have the advantage that they can be wrapped around the beam pipe if deemed necessary after the accelerator has been constructed, e.g. this was done at Fermilab~\cite{fermilab1, fermilab2}. Alternatively, a mu-metal layer could be incorporated into the beam pipe design and the entire beam pipe could be annealed in its final form, which would ensure a good shielding performance.

\subsection{Magnets}
Beam pipes are typically formed from non-magnetic materials because they run through the aperture of magnets. They should not impede the magnetic field generated by a magnet, which is used to guide the beam.

The sensitivity to stray magnetic fields in linear colliders comes from the long drifts between magnets. Therefore, only the drifts need to be shielded, which avoids the problem of shielding inside the magnets.

Large static magnetic fields saturate ferromagnetic materials. Once a material is saturated, it is no longer effective as a magnetic shield. Depending on the required internal field level, this property enables the possibility of replacing the steel in a beam pipe with soft iron. Inside a magnet the soft iron beam pipe will be saturated and will not impede the magnetic field, whereas in the drifts the soft iron beam pipe will shield the beam.

\subsection{Factors Affecting Performance}
Factors that affect the performance of magnetic shields are discussed in~\cite{magnetic-shielding}. The factors that affect the shielding performance of dynamic magnetic fields are summarised below.

\subsubsection{Saturation}
Eq.\,\eqref{e:rayleigh-permeability} is valid provided a static magnetic field does not saturate the material. if the material is saturated, its permeability and shielding performance drops. Using Eq.\,\eqref{e:mu-metal-simple-tf} it is straightforward to show a mu-metal shield will not saturate provided
\begin{equation}
B_s > \frac{D}{\Delta} H,
\end{equation}
where $B_s$ is the magnetic induction at saturation. The magnetic induction for the mu-metal used in this work is $B_s=0.74$\,T (see Table~\ref{t:material-specifications}). The dominant static magnetic field in an accelerator environment is typically the Earth's magnetic field, which is approximately 20-70\,$\mu$T~\cite{earth-field}. Assuming 50\,$\mu$T for the Earth's magnetic field, this requires a shield geometry that satisfies $D/\Delta < 15,000$, which is easily achieved. The magnetic shields considered in this work have a $D/\Delta$ between 10 and 1000.

Alternatively, an additional outer layer can be included in the shield, which has a higher magnetic induction at saturation, e.g. a nickel-iron alloy with a lower nickel content than mu-metal~\cite{ni-fe-alloys}. The outer layer will attenuate the static magnetic field and ensure an inner mu-metal layer does not saturate.

\subsubsection{Annealing}
Soft ferromagnetic materials are often annealed in a dry hydrogen environment after being bent into their final form. This removes impurities from the material and alters the crystal structure of the material, which allows  magnetic domains to move freely \cite{bozorth, annealing1}. As a result, the permeability of the material is significantly increased~\cite{annealing2, annealing3, annealing4}.

\subsubsection{Mechanical Stress, Deformation and Shock} \label{s:performance-stress}
It is well known that mechanical stress, deformation and shock can significantly reduce the permeability of a ferromagnetic material~\cite{stress1, stress2}. The damage can be reversed by re-annealing the shield, which can increase the permeability by an order of magnitude~\cite{annealing4}. Mu-metal requires hydrogen annealing at very high temperatures (above 1000$^\circ$C~\cite{annealing4}) which means re-annealing in the accelerator tunnel impractical. The sample should be handled with care after annealing to avoid performance loss.

\subsubsection{Temperature}
It was observed in~\cite{kek2, annealing4} that the shielding factor of a mu-metal shield degrades at very low (superconducting) temperatures. This is only a concern for accelerators that operate at superconducting temperatures, such as the ILC. CLIC operates at room temperature, which means the degradation of shielding at low temperatures is not a concern.

\section{Conclusions}
The behaviour of the permeability for very small-amplitude magnetic fields (Rayleigh's law) has been verified. It is possible to shield extremely small-amplitude magnetic fields, down to the level of 0.1\,nT, with mu-metal. Mu-metal is sensitive to permeability loss from mechanical stress and deformation. It should be handled with care after annealing. A simple formula (Eq.\,\eqref{e:mu-metal-simple-tf}) was verified for calculating the transfer function of mu-metal.

There is an increasing need to shield beams in accelerators from external magnetic fields, in particular for future linear colliders. In this paper, we have confirmed experimentally that mu-metal is a viable material that can be used to shield dynamic magnetic fields to amplitudes of less than 0.1\,nT. This is particularly important for CLIC which requires the stray magnetic fields experienced by the beam do not exceed 0.1\,nT. A mu-metal shield has been included in the design of CLIC for this purpose~\cite{thesis}. A discussion of advantages and disadvantages of other mitigation systems in CLIC can be found in~\cite{thesis}, however the mu-metal shield is the preferred option.

\end{document}